\documentclass[amsmath,amssymb,superscriptaddress,nobalancelastpage,prb,twocolumn]{revtex4-1}

\usepackage{graphicx}
\usepackage{varioref}
\usepackage{xr-hyper}
\usepackage{xcolor}
\usepackage{nicefrac}
\usepackage{xfrac}
\usepackage{hyperref}
\hypersetup{colorlinks,linkcolor=blue,urlcolor=blue,citecolor=orange}
\usepackage{ulem}
\usepackage{siunitx}
\usepackage{graphicx}% Include figure files
\usepackage{dcolumn}% Align table columns on decimal point
\usepackage{bm}% bold math
\usepackage{braket}
\usepackage{wasysym}
\usepackage{textcomp}

\newcommand{\lsco}{{La}$_{1.88}${Sr}$_{0.12}${CuO}$_4$}
\newcommand{\lscoF}{{La}$_{2-x}${Sr}$_{x}${CuO}$_4$}

\begin{document}

\title{Single-domain stripe order in a high-temperature superconductor}

\author{G.~Simutis}
\email{gediminas.simutis@psi.ch}
\affiliation{Laboratory for Neutron and Muon Instrumentation,
Paul Scherrer Institut, CH-5232 Villigen PSI, Switzerland}
\affiliation{Department of Physics, Chalmers University of Technology, SE-41296 G\"{o}teborg, Sweden}

\author{J.~K\"uspert}
\affiliation{Physik-Institut, Universit\"{a}t Z\"{u}rich, Winterthurerstrasse 190, CH-8057 Z\"{u}rich, Switzerland}

\author{Q.~Wang}
\affiliation{Physik-Institut, Universit\"{a}t Z\"{u}rich, Winterthurerstrasse 190, CH-8057 Z\"{u}rich, Switzerland}

\author{J.~Choi}
\affiliation{Diamond Light Source, Harwell Campus, Didcot OX11 0DE, United Kingdom}

\author{D.~Bucher}
\affiliation{Physik-Institut, Universit\"{a}t Z\"{u}rich, Winterthurerstrasse 190, CH-8057 Z\"{u}rich, Switzerland}

\author{M.~Boehm}
\affiliation{Institut Laue-Langevin,71 avenue des Martyrs, CS 20156, 38042 Grenoble Cedex 9, France}

\author{F.~Bourdarot}
\affiliation{Universit\'{e} Grenoble Alpes, CEA, INAC, MEM MDN, 38000 Grenoble, France}

\author{M.~Bertelsen}
\affiliation{European Spallation Source ERIC, P.O. Box 176, SE-221 00, Lund, Sweden}

\author{C. N.~Wang}
\affiliation{Laboratory for Muon Spin Spectroscopy, Paul Scherrer Institut, CH-5232 Villigen PSI, Switzerland}

\author{T.~Kurosawa}
\affiliation{Department of Physics, Hokkaido University - Sapporo, 060-0810 Sapporo, Hokkaido, Japan}

\author{N.~Momono}
\affiliation{Department of Physics, Hokkaido University - Sapporo, 060-0810 Sapporo, Hokkaido, Japan}
\affiliation{Department of Applied Sciences, Muroran Institute of Technology, Muroran 050-8585, Japan}

\author{M.~Oda}
\affiliation{Department of Physics, Hokkaido University - Sapporo, 060-0810 Sapporo, Hokkaido, Japan}

\author{M.~M{\aa}nsson}
\affiliation{Department of Applied Physics, KTH Royal Institute of Technology, SE-106 91 Stockholm, Sweden}

\author{Y.~Sassa}
\affiliation{Department of Physics, Chalmers University of Technology, SE-41296 G\"{o}teborg, Sweden}

\author{M.~Janoschek}
\affiliation{Laboratory for Neutron and Muon Instrumentation,
Paul Scherrer Institut, CH-5232 Villigen PSI, Switzerland}
\affiliation{Physik-Institut, Universit\"{a}t Z\"{u}rich, Winterthurerstrasse 190, CH-8057 Z\"{u}rich, Switzerland}

\author{N.~B.~Christensen}
\affiliation{Department of Physics, Technical University of Denmark, DK-2800 Kongens Lyngby, Denmark}

\author{J.~Chang}
\email{johan.chang@physik.uzh.ch}
\affiliation{Physik-Institut, Universit\"{a}t Z\"{u}rich, Winterthurerstrasse 190, CH-8057 Z\"{u}rich, Switzerland}

\author{D.~G.~Mazzone}
\email{daniel.mazzone@psi.ch}
\affiliation{Laboratory for Neutron Scattering and Imaging,
Paul Scherrer Institut, CH-5232 Villigen PSI, Switzerland}

\date{\today}% It is always \today, today,
             
\begin{abstract}
\end{abstract}

\maketitle

\textbf{The coupling of spin, charge and lattice degrees of freedom results in the emergence of novel states of matter across many classes of strongly correlated electron materials. A model example is unconventional superconductivity, which is widely believed to arise from the coupling of electrons via spin excitations. In cuprate
high-temperature superconductors, the interplay of charge and spin degrees of freedom is also reflected in a zoo of charge and spin-density wave orders that are intertwined with superconductivity. A key question is whether the different types of density waves merely coexist or are indeed directly coupled. Here we use a novel neutron diffraction technique with superior beam-focusing that allows us to probe the subtle spin-density wave order in the prototypical high-temperature superconductor \lsco~under applied uniaxial pressure to demonstrate that it is immediately coupled with charge-density wave order. Our result shows that suitable models for high-temperature superconductivity must equally account for charge and spin degrees of freedom via uniaxial charge-spin stripe fluctuations.}

A universal question in strongly correlated electron materials is how the salient degrees of freedom are microscopically coupled to stabilize novel quantum states.
\begin{figure*}[tbh]
\centering
\includegraphics[width={\textwidth}]{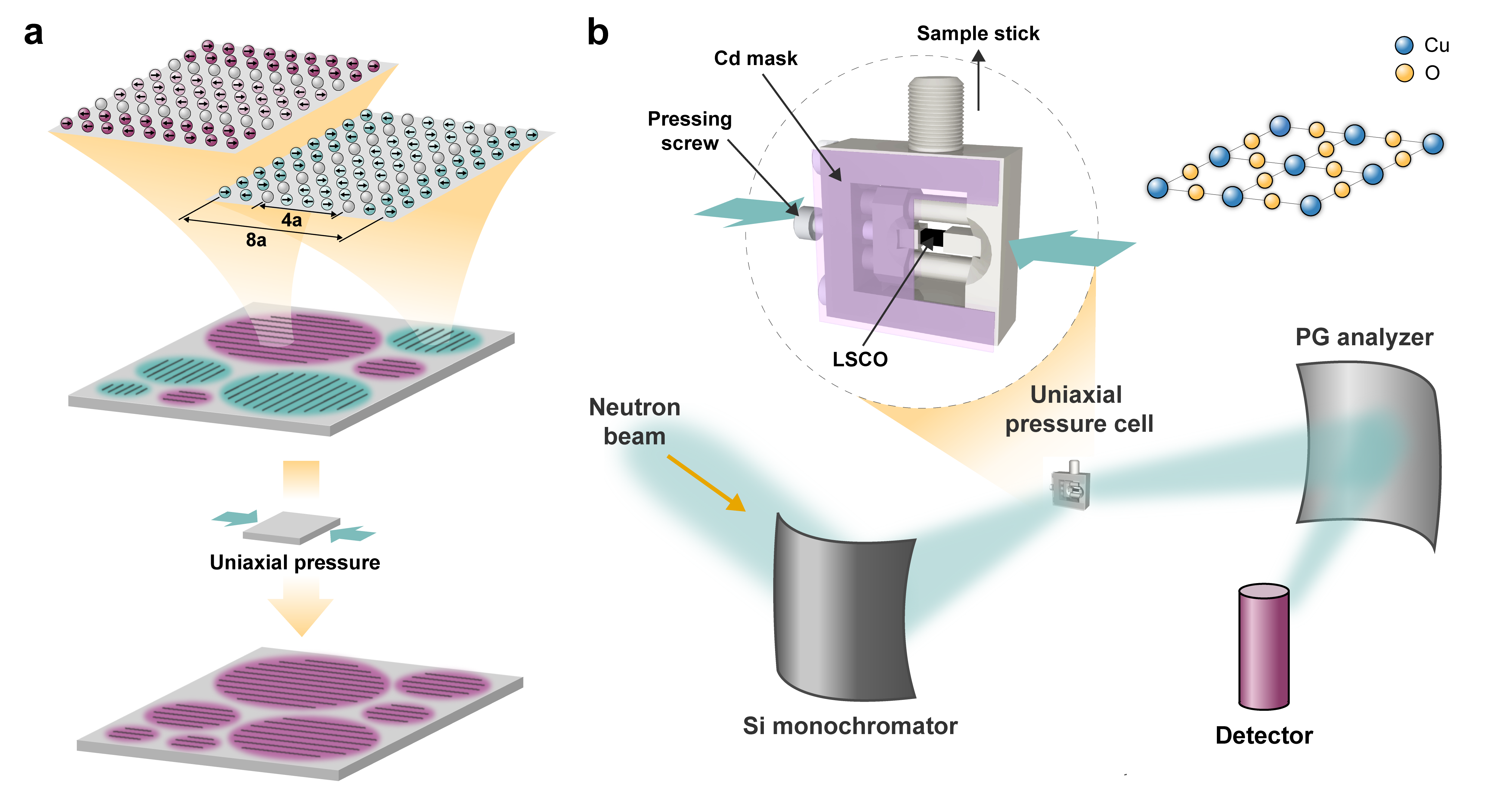}
\caption{\label{fig:setup}%
\textbf{Uniaxial pressure tuning of the LSCO electronic ground state.} \textbf{a} The low-temperature ambient pressure ground state of \lsco~is thought to possess equally populated CDW and SDW domains with $\delta_\text{CDW}$ = 2$\delta_\text{SDW}$, where $\delta_\text{CDW}$ and $\delta_\text{SDW}$ are the incommensurations of the CDW and SDW wavevectors reflected in the 4$a$ and 8$a$ periodicity of the density waves in direct space, respectively (see also Methods section). If so, the application of uniaxial pressure should lift the ground state degeneracy, clarifying whether spin and charge degrees of freedom are directly coupled in a single-domain stripe state.  \textbf{b} Setup used in our neutron diffraction experiment. A  cuboid shaped 55 mg \lsco~single crystal was mounted into a uniaxial pressure cell, and strained along the Cu-O bond direction. The specially designed pressure cell was made of pure aluminium and masked with neutron absorbing cadmium to minimize background contributions. The static SDW order was studied on the cold triple-axis spectrometer ThALES at the Institut Laue-Langevin, using a double focusing silicon monochromator and pyrolytic graphite (PG) analyzer to efficiently collect neutrons diffracted from the SDW order in a $^3$He detector.}
\end{figure*}
Unconventional superconductivity in cuprate materials is a model example, where it is thought that the macroscopic coherent quantum state arises from intertwined charge (CDW) and spin-density wave (SDW) fluctuations~\cite{Tranquada2013,Tranquada2021, WenNatComm2019,Jiang2019,Mingpu2022}. A variety of experimental results alongside theoretical Hubbard model calculations show that the ground state energies of static CDW,  SDW orders and unconventional superconductivity are nearly degenerate~\cite{Tranquada2013,Tranquada2021,KivelsonRMP2003,HuangSci2017,ZhengSci2017,Corboz2014}. However, it remains unclear how charge and spin degrees of freedom are coupled to enable high-temperature superconductivity in distinct cuprate materials.

An effective way to tune the subtle balance among the different ground states is the inclusion of holes into the copper-oxygen layers, which has  notably established that the density-wave orders are most stable when the number of holes per Cu site $p$ = 1/8. Interestingly, this is observed concomitant with a suppression of superconductivity. In addition, in YBa$_2$Cu$_3$O$_{7-y}$, for instance, static CDW and SDW orders seem not to coexist~\cite{BlancoCanosaPRL2013,HuckerPRB2014}. This is in striking contrast with La-based cuprates where the static density-wave orders not only coexist but also show signatures of coupling~\cite{Tranquada2013,Tranquada2021,Corboz2014}.
Yet the two order parameters reveal very different dependencies on temperature and hole-doping. 
In \lscoF~(LSCO) with $x$ = $p$ = 1/8 the SDW order is strongest revealing an onset temperature comparable to that of superconductivity, $i.e.$ $T_c \approx$ 27~K~\cite{ChangPRB2008}, but short-range CDW order persists to temperatures above 100~K \cite{WenNatComm2019}. 
Similarly, while density-wave fluctuations exist over a broad hole-doping range in LSCO \cite{WenNatComm2019,Dean2013},  SDW and CDW orders coexist only for $\sim$0.1 $<x<$ $\sim$0.135 questioning whether they vanish in a common quantum critical point~\cite{MiaonpjQM2021,WenNatComm2019,MaPRR2021,FrachetNatPhys2020,LinPRL2020,gupta_vanishing_2021}. Summarizing, this raises the question how SDW and CDW orders are microscopically coupled, and how their fluctuations give rise to unconventional superconductivity.

To address this issue for \lsco, we leverage uniaxial pressure applied along the copper-oxygen (Cu-O) bond direction as a newly-established extrinsic tuning parameter. Here recent x-ray experiments~\cite{Choi20} have demonstrated that one of the CDW domains can be suppressed through uniaxial strain.
For the case of truly coupled CDW and SDW orders, we similarly expect to observe that uniaxial pressure generates a single domain stripe SDW order (see Fig.~\ref{fig:setup}). Here the outstanding experimental challenge is that sufficiently high strains may only be achieved for tiny single crystals with well-defined shapes and small surface cross-sections perpendicular to the direction of the applied pressure. Such crystals of a few mg weight are typically not suitable for neutron diffraction experiments required to unambiguously probe weak SDW order. Crucially, the scattered neutron intensity scales linearly with the crystal mass while uniaxial pressure cells generally increase background scattering, resulting in an overall insufficient signal-to-noise ratio.

To overcome this challenge, we designed a neutron study that  combines three major technical advances. Firstly, we exploit new developments in neutron-ray-tracing simulations, where the Union components of the McStas suite~\cite{Bertelsen2019} only since recently take into account background scattering processes resulting from advanced sample environments such as cryostats and pressure cells. This, in turn, allowed us to optimize the signal-to-background ratio (see Supplementary Information (SI) section II) and demonstrate that irrespective of geometrical and instrumental details, neutrons at longer wavelengths are advantageous in terms of signal-to-noise ratio. Secondly, armed with this knowledge, we leveraged the unique focusing capabilities of the adaptive virtual source and double focusing Si(111) monochromator at the ThALES cold neutron triple-axis spectrometer at the Institut Laue-Langevin, Grenoble, France to probe the weak SDW order in a 55~mg LSCO $x$ = 0.12 single crystal~\cite{Boehm_2008,Boehm_2015, Thales2013}. Finally, this strongly reduced sample size (for comparison, a 1.06~g sample from the same batch was used for ambient pressure measurements. See Methods section), enabled the use of a scaled-up version of a uniaxial pressure cell, which we recently developed for x-ray studies of CDW order~\cite{Choi20,Wang22},  without compromising uniaxial pressure application  (see Fig. \ref{fig:setup}b). We note that the exquisite focusing capability of ThALES not only strongly increased the flux on the tiny sample but equally important reduced background scattering, allowing for a vastly improved signal-to-noise ratio (further details are given in the Methods section and in the SI section III).

\begin{figure}[tbh]
\centering
\includegraphics[width={\linewidth}]{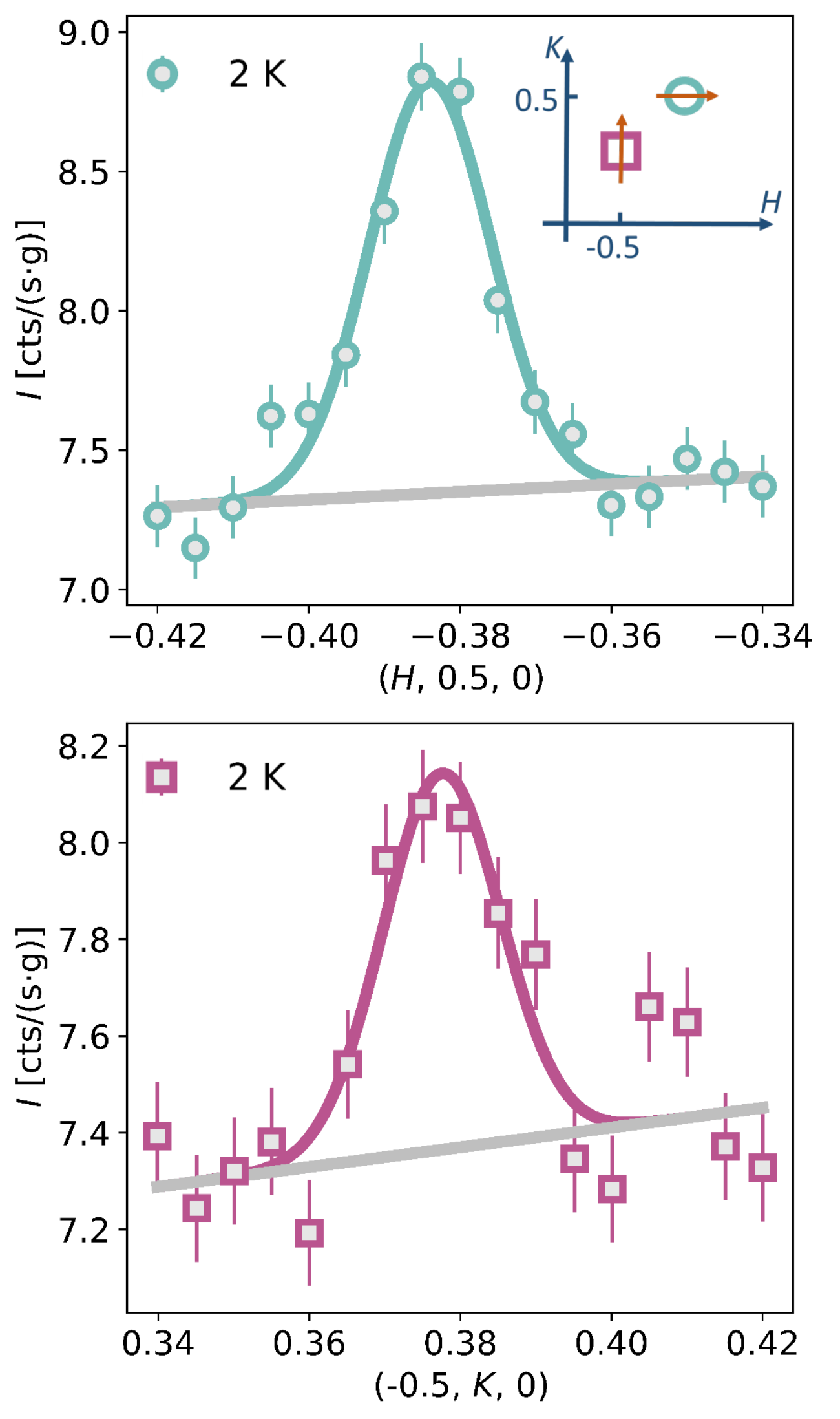}\\
\caption{\label{fig:large}%
\textbf{Magnetic intensity under zero strain conditions.} The figure displays magnetic neutron diffraction intensity along the two reciprocal space directions indicated in the inset. The \lsco~sample was measured at ambient pressure and $T$ = 2~K. Intensities ($I$) are given in counts (cts) per seconds (s) and mass (g) of the sample. Error bars are dictated by Poisson statistics. Scans through \textbf{Q}$^\text{a}_\text{SDW}$ (circles) and \textbf{Q}$^\text{b}_\text{SDW}$ (squares) reveal magnetic intensity at both wavevectors. The data were fitted with a Gaussian line shape on a sloping background.}
\end{figure}

\begin{figure*}[tbh]
\centering
\includegraphics[width={\textwidth}]{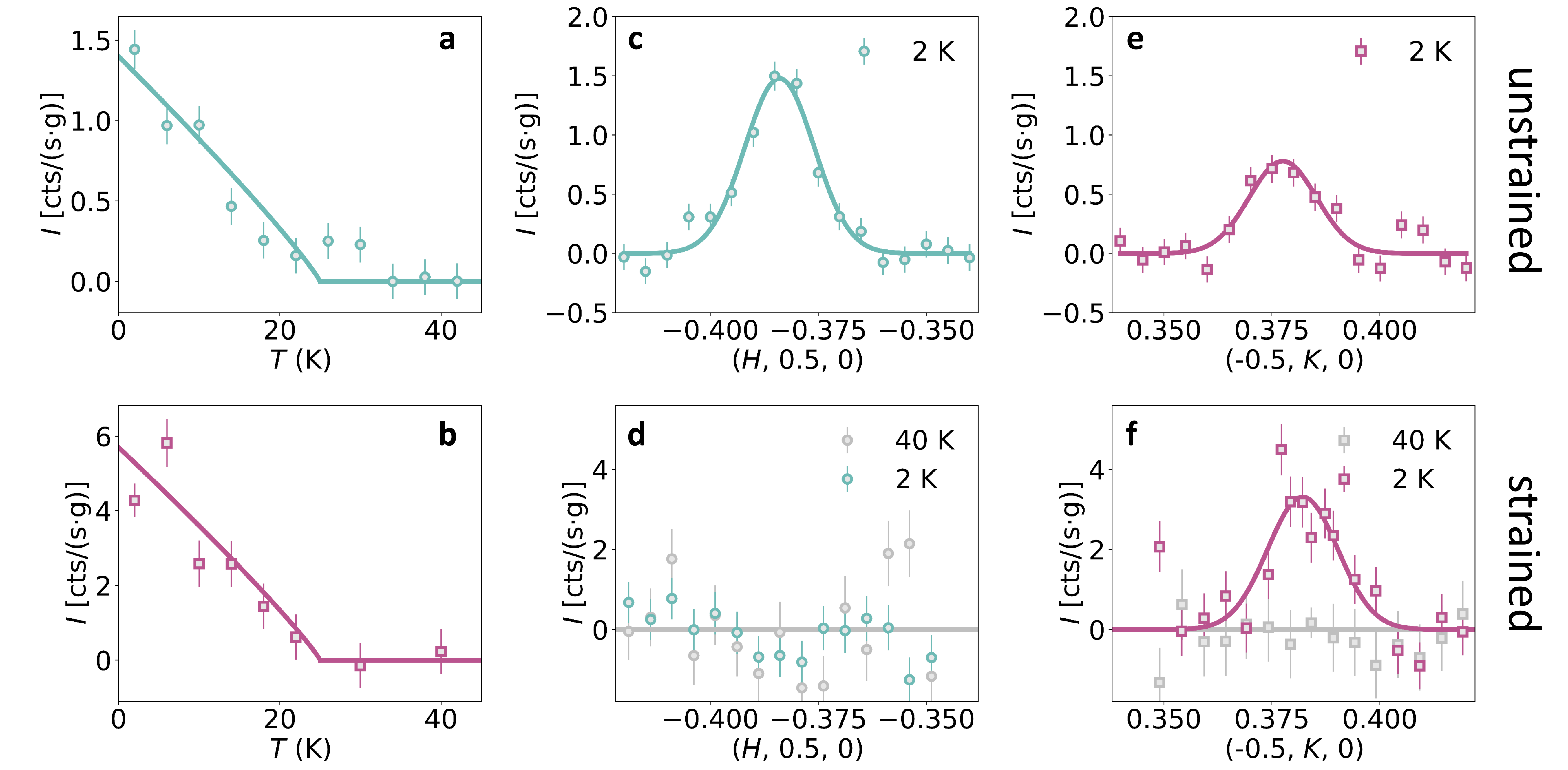}\\
\caption{\label{fig:peaks}%
\textbf{Strain-induced SDW-domain repopulation.}
Background-subtracted neutron scattering results comparing the scattering from the unstrained (\textbf{a, c, e}) and strained (\textbf{b, d, f}) sample. Panels \textbf{a} and \textbf{b} show the temperature dependence of the magnetic Bragg peak intensities at \textbf{Q}$^\text{a}_\text{SDW}$ and \textbf{Q}$^\text{b}_\text{SDW}$ for the unstrained and stained sample, respectively. The solid line is a guide to the eye. Panels \textbf{c} and \textbf{d} display the reciprocal space scans of \lsco~around \textbf{Q}$^\text{a}_\text{SDW}$ at ambient conditions and under compressive $a$-axis strain. In panels \textbf{e} and \textbf{f} we plot the analogous scans around \textbf{Q}$^\text{b}_\text{SDW}$. Their comparison shows that uniaxial pressure along the LSCO Cu-O bond direction leads to a repopulation of the SDW domains, in which the magnetic domain perpendicular to the pressing direction is favored.}
\end{figure*}

Figure~\ref{fig:large} shows the peaks characteristic of the SDW order in \lsco~at ambient pressure and a temperature $T$ = 2~K (additional scans are plotted in SI section V) as probed by our experiment on the 1.06~g sample. The measured count rates are shown normalized to the sample mass to facilitate a later comparison with results obtained with the pressure cell. Within the high-temperature tetragonal unit cell notation SDW order in La-based cuprates manifests at incommensurate wavevectors \textbf{Q}$^\text{a}_\text{SDW}$ = (0.5 $\pm$~$\delta_\text{SDW}$, 0.5, 0) and \textbf{Q}$^\text{b}_\text{SDW}$ = (0.5, 0.5 $\pm$~$\delta_\text{SDW}$, 0) in reciprocal lattice units (rlu). They may be interpreted as the signature of either two orthogonal magnetic domains in the well-known stripe model~\cite{TranquadaNat1995, Tranquada2013, Tranquada2021, Fujita2004, Berg_2009,Abbamonte2005, Berg_2009,Seibold2007, KivelsonRMP2003} shown in Fig.~\ref{fig:setup}a, or to two phase-related wavevectors describing a single-domain multi-\textbf{Q} SDW order, which is consistent with the interpretation of earlier experimental results~\cite{Christensen2007, Fine2007, Fine2011, Robertson2006, Zachar1998, Aristova2019, Abbamonte2005, Seibold2007, Wise2008, Dolgirev2017, Wang_2019, Brandenburg2013}. The incommensurability $\delta_\text{SDW}$ is doping dependent with $\delta_\text{SDW}\approx 0.12$~rlu around $x$ = 1/8 \cite{Tranquada2013}. Our measurements unambiguously show that magnetic intensity is present at both wavevectors \textbf{Q}$^\text{a}_\text{SDW}$ and \textbf{Q}$^\text{b}_\text{SDW}$. The data were fitted with a Gaussian line shape on a sloping background, revealing an average incommensuration $\delta_\text{SDW}$ = 0.118(4)~rlu and an in-plane correlation length $\xi_a$ = $(a$/$\pi)$FWHM$^{-1}$ = 67(9)~\AA, where FWHM is the full-width at half-maximum. The temperature dependence of the intensity at \textbf{Q}$^\text{a}_\text{SDW}$ is shown in Fig.~\ref{fig:peaks}a, providing evidence that magnetic long-range order is suppressed above $T_N$ $\approx$ 25~K.  All these results are in agreement with earlier reports on \lsco~\cite{He21,ChangPRB2008}. We also observe a small perpendicular incommensuration (0.007(2)~rlu shown in SI section V), which is often referred to as $Y$-shift and is hypothesised to arise through pinning to orthorhombic distortions \cite{He21, KimuraPRB2000, Jacobsen2015}.

\begin{figure}[tbh]
\centering
\includegraphics[width={\linewidth}]{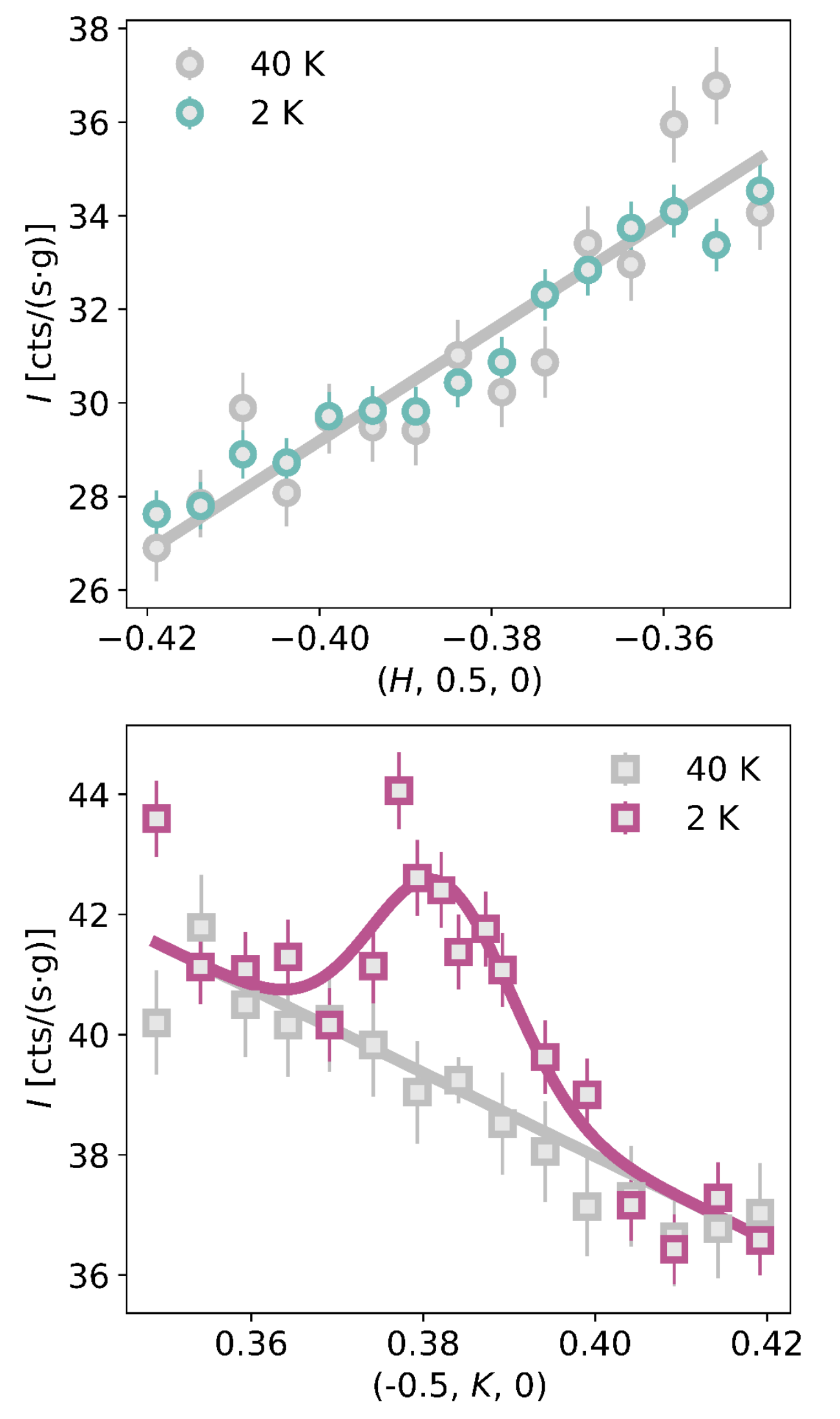}\\
\caption{\label{fig:small}%
\textbf{Single magnetic domain state under uniaxial pressure.}
Reciprocal space scans of \lsco~under compressive $a$-axis strain measured and $T$ = 2 and 40 K. Intensities ($I$) are given in counts (cts) per seconds (s) and mass (g) of the sample. Error bars are dictated by Poisson statistics. Scans along \textbf{Q}$^\text{a}_\text{SDW}$ and \textbf{Q}$^\text{b}_\text{SDW}$ reveal a single domain state. The high-temperature data were fitted with a sloping background. The low-temperature data of the \textbf{Q}$^\text{b}_\text{SDW}$-domain were fitted to a Gaussian line shape over the fixed high-temperature background.
}
\end{figure}

To probe the response of SDW order in \lsco~to uniaxial pressure, we applied compressive strain $\epsilon_a$ $\approx$ 0.02\% (see Methods section) along the $a$-axis (Cu-O bond direction) of the 55~mg \lsco~sample (cf. Fig.~\ref{fig:setup}).  In Fig.~\ref{fig:small}, we display reciprocal space scans across the two wavevectors at $T$ = 2 and 40~K. The high temperature data were measured above the SDW onset temperature (see Fig.~\ref{fig:peaks}a and b), and used as background calibration assuming a sloping behavior. The high counting statistics enabled by our setup, allowed us to observe a magnetic signal down to approximately 2\% above the background (see SI section VI). At this level of discrimination no magnetic Bragg peak is observed at $T$ = 2~K around (-0.5 + $\delta_\text{SDW}$, 0.5, 0) or other symmetry equivalent reflections of \textbf{Q}$^\text{a}_\text{SDW}$ (see SI section VI). In strong contrast, a clear magnetic signal was found at \textbf{Q}$^\text{b}_\text{SDW}$ (see Fig.~\ref{fig:small} and SI section VI). 
We fitted the roughly 8\% strong signal at (-0.5, 0.5 - $\delta_\text{SDW}$, 0) with a Gaussian lineshape over the fixed high-temperature background. The fit yields $\delta_\text{SDW}$ = 0.118(2)~rlu and $\xi_b$ = 66(13)~\AA, matching the ambient pressure results. 
We note that a slight 0.005(1)~rlu Y-shift persists under strain (see SI section VI) and that no evidence for a pressure-induced shift in $T_N$ is detected (see Fig.~\ref{fig:peaks}a and b). 

For a quantitative comparison of the data obtained at ambient conditions and under $a$-axis strain, the background contributions from the scans shown in Fig.~\ref{fig:large} and \ref{fig:small} were substracted and are displayed in Fig.~\ref{fig:peaks}c-f. 
The peaks associated with \textbf{Q}$^\text{a}_\text{SDW}$ and \textbf{Q}$^\text{b}_\text{SDW}$ feature integrated intensities $I_\text{int}$ = 0.030(3) and 0.015(3)~cts$\cdot$rlu/(s$\cdot$g) at ambient pressure, respectively. By contrast, under uniaxial strain conditions only \textbf{Q}$^\text{b}_\text{SDW}$ carries a finite integrated intensity of $I_\text{int}$ = 0.07(1)~cts$\cdot$rlu/(s$\cdot$g). Alongside our finding that strain has no effect on the SDW ordering vector, correlation length or onset temperature, this demonstrates that uniaxial pressure along the Cu-O bond direction leads to a dramatic redistribution of magnetic peak intensity. This observation allows us to gain insight into the SDW structure. While extrinsic tuning parameters may modify the population of magnetic domains by breaking the symmetry on a macroscopic scale, they are not expected to differently affect the coexisting wavevectors in multi-\textbf{Q} structures~\cite{RODRIGUEZCARVAJAL2019770,Wills2001}. 
Thus, our results provide direct experimental evidence for a single-$\bf Q$ magnetic structure in LSCO, and that application of pressure along one of the Cu-O directions favors the domain characterized by a propagation vector along the perpendicular Cu-O direction. We note that this effectively rules out spin-vortex checkerboard structures and any other multi-\textbf{Q} single-domain structures proposed earlier~\cite{Christensen2007, Fine2007, Fine2011, Robertson2006, Aristova2019, Seibold2007, Dolgirev2017, Brandenburg2013}.

Our observation that the application of uniaxial strain causes domain repopulation reflected in the peak intensity without altering the other characteristics of the magnetic order, enables us to firmly establish the link between SDW and CDW order in \lsco. Recent x-ray scattering results have shown that small uniaxial strain along the Cu-O bond direction also yields a suppression of the intensity of the CDW ordering vector along the direction of the applied pressure. Together with our neutron diffraction study these results strongly suggests a direct coupling between CDW and SDW order. This provides clear-cut evidence for a charge-spin stripe arrangement (see Fig.~\ref{fig:setup}a) as the fundamental density-wave state of LSCO. A direct coupling between static charge and spin order is supported by a microscopic mechanism in the strong coupling limit, where stripe order arises from local correlations \cite{MiaoPRX2019, Tranquada2013, MiaoPRX2018, Edwin2017, Zheng2017}. In this picture holes are located at the antiphase SDW domain boundaries (see Fig.~\ref{fig:setup}a) to minimize their kinetic energy, therefore establishing the relationship $\delta_\text{CDW}$ = 2$\delta_\text{SDW}$ between the incommensurations of the two orders. As the direct space picture naturally connects charge and spin order, uniaxial pressure must induce a single domain state where populated charge and spin domains are oriented along the same direction. Our observations are harder to reconcile with a weak coupling picture where the electronic order arises from Fermi surface nesting instabilities \cite{gruener1988, gruener1994}. In this case, uniaxial pressure affects the Fermi surface topology, which would need to act similarly at $\textbf{Q}_\text{SDW}$ and $\textbf{Q}_\text{CDW}$ to account for a unidirectional charge and spin domain states. To explain the observation that only one ordering vector survives the pressure application, significant and peculiar anisotropic Fermi-surface distortions would be required. In contrast to our observations, one would also expect subtle shifts of the ordering wavevectors in such a scenario.

Our observation of single-\textbf{Q} charge-spin stripe order in La-based cuprates adds to the discussion on the interplay between density-wave orders and unconventional superconductivity. Several theories attest the intertwined coupling between superconductivity and density-wave order to the emergence of a spatially-modulated superconducting order parameter that makes up for a sizable fraction of the superconducting condensate \cite{Tranquada2013, WenNatComm2019, Tranquada2021}. This putative pair-density wave (PDW) phase is thought to change sign within the spin stripes to cope with magnetic long-range order. Because both the SDW order probed here, and the CDW order explored by our previous x-ray scattering study~\cite{Choi20} is not altered by uniaxial pressure along the Cu-O bond direction, this suggests that any putative PDW order will also consist of uniaxial stripes. Notably, that the magnetic correlation length and SDW onset temperature, as well as the coupling between CDW order and superconductivity remain unchanged by uniaxial pressure application, suggests that in LSCO charge-spin stripe order and unconventional superconductivity are deeply intertwined. Interestingly, for {La}$_{2-x}${Ba}$_{x}${CuO}$_4$ (LBCO) a different behavior has been observed when uniaxial pressure is applied along an in-plane axis close to the Cu-Cu bond direction, demonstrating direct competition between superconductivity and magnetic order\cite{Guguchia2020,Kamminga2022}. Thus, further uniaxial pressure studies on LSCO and LBCO along different crystal directions will be crucial to gain new insight into the coupling between stripe order and unconventional superconductivity. 

In addition, we note that despite these differences, short-ranged charge fluctuations generally exist over a wider temperature and doping range than spin order in La-based cuprates~\cite{MiaonpjQM2021,WenNatComm2019,MaPRR2021,FrachetNatPhys2020,LinPRL2020,gupta_vanishing_2021}. This suggests that the charge degrees of freedom are responsible for the primary fluctuations that drive the coupled charge-spin state at low temperature. Our results add further constraints to the intertwined ground state of this strongly correlated material class. They attest that  adequate theories for high-temperature superconductivity must account for charge and spin degrees of freedom via uniaxial charge-spin fluctuations. Future experiments directly assessing the spin fluctuations under uniaxial pressure may be a fruitful way forward to gain a deeper understanding of the pairing mechanism behind high-temperature superconductivity.

Beyond the cuprates, the exact nature of the intimate coupling among various electronic degrees of freedom is crucial to the understanding of all strongly-correlated electron materials~\cite{Paschen2020,Dagotto2005,Fiebig2016}. Our study establishes that the latest-generation neutron instrumentation allows us to explore the effects of uniaxial pressure --- a particularly clean and useful tuning parameter --- on spin order with weak order parameters. This advancement will be key in improving our understanding of a large range of electronic ground states. Notably, the study of coupled PDW and SDW order in heavy-fermion superconductors~\cite{Gerber2014,Kim2016}, coupling of ferroelectric and magnetic domains in multiferroics~\cite{Fiebig2016}, spin, charge and lattice coupling in electronic nematic states~\cite{Ronning2017,Fobes2018,Fernandes2014}, or the identification of topologically non-trivial multi-\textbf{Q} spin textures\cite{Fert2017}~will benefit from this capability. 

\section*{Methods}
\textbf{Sample preparation and characterization.}
A high quality LSCO $x$ = 0.12 crystal was grown by the travelling solvent floating zone technique. Standard magnetization measurements revealed a superconducting transition temperature $T_c$ = 27~K  (cf. SI section I). The magnetic response was checked using muon spin rotation (MuSR) at the GPS beamline of the S$\mu$S muon source, Paul Scherrer Institut, Switzerland (see SI section I). These measurements showed that the magnetically ordered state inside the superconducting phase is in line with the earlier studies on LSCO \cite{Savici2002,ChangPRB2008}.

After confirming the rod's surface crystallinity by x-ray Laue diffraction, two samples of masses $m$ = 1.060 and 0.055~g were oriented and cut using a tungsten wire saw.
The latter sample dedicated to fit into the uniaxial pressure cell had a cuboid shape of dimensions 3.17(5)x1.35(5)x1.83(5) mm$^3$ along the axis of the high-temperature tetragonal unit cell. The tetragonal notation (see SI section I for details) with $a$ = $b$ = 3.759~\AA~ and $c$ = 13.2~\AA~ refined at $T$ = 2~K on ThALES is employed throughout the manuscript. Here the SDW wavevector $\textbf{Q}_\text{SDW}$ is shifted by $\textbf{q}_\text{SDW}$ = ($\delta_\text{SDW}$, 0, 0)  and (0, $\delta_\text{SDW}$, 0) away from the antiferromagnetic wavevector $\textbf{Q}_\text{AF}$ = (1/2, 1/2, 0). The charge order peaks at $\textbf{Q}_\text{CDW}$ are offset by $\textbf{q}_\text{CDW}$ = (2$\delta_\text{SDW}$, 0, 0) and (0, 2$\delta_\text{SDW}$, 0) with respect to structural Bragg peaks, yielding $\delta_\text{CDW}$ = 2$\delta_\text{SDW}$.

The bulk crystallinity of the two samples was checked using the triple-axis alignment station IN3, the neutron Laue diffractometer Orient Express and finally determined on the cold neutron spectrometer ThALES at the Institut Laue-Langevin, Grenoble, France. These data revealed that the small 55~mg sample was single crystalline within our resolution. The larger 1.06 g was composed of two crystallites shifted by 1.5 degrees with a relative intensity of 5:1 that was determined at $T$ = 2~K (see SI section IV).

A strain cell similar to those used in recent x-ray diffraction studies \cite{Choi20,Wang22} was adapted to fit the cryostat dimensions of the ThALES instrument at the Laue-Langevin (see also Fig. \ref{fig:setup}b). The cell was constructed from high purity aluminium to minimize background contributions. Uniaxial pressure was applied at room temperature along the $a$-axis using an M3 aluminium screw that was rotated by 60$^\circ$. Following the calibration from Ref. \cite{Choi20} this amounts to a compressive strain of $\epsilon_a$ = $\Delta a/a$ $\approx$ 0.02\%.

\textbf{Neutron Diffraction at ThALES.} The neutron scattering experiment was performed at the high-flux cold neutron spectrometer ThALES at the Institut Laue-Langevin. ThALES is uniquely suited for the purpose of our experiment since its double-focusing Si(111) monochromator allows aggressive focusing of the incident neutron at the sample position \cite{Boehm_2008,Boehm_2015, Thales2013}. This was essential to minimize parasitic scattering from the strain cell while maximizing the signal from the sample. A sketch of the experimental setup is shown in Fig.~\ref{fig:setup}b. We used $k_i=k_f=1.55$~\AA$^{-1}$ neutrons, which were cleaned from higher order contaminations via a velocity selector before the silicon monochromator  
and a cooled Be-filter before a double-focusing PG(002) analyser. The instrument was used in a $W$-configuration  to minimize the extent of the instrumental resolution ellipsoid. The scattered neutrons were detected with a standard one-inch $^3$He detector. The samples were aligned perpendicular to the (0, 0, 1) axis to access both SDW domains in the horizontal scattering plane. We improved the signal-to-noise ratio for the strain sample by optimizing the position and opening (8~mm) of the virtual slit, which is located between the velocity selector and the monochromator (see SI section III). Further improvements of the signal-to-noise ratio were gained through tight slit openings before (8~mm horizontal, 15~mm vertical) and after (10.5~mm horizontal, 13~mm vertical) the sample.

The temperature dependencies along \textbf{Q}$^\text{a}_\text{SDW}$ and \textbf{Q}$^\text{b}_\text{SDW}$ shown in Fig.~\ref{fig:peaks}a and b were measured on the optimized peak positions, i.e. at $\textbf{Q}$ = (-0.387, 0.512, 0) and (-0.504, 0.382, 0), respectively. Additional reciprocal lattice scans on the 1.06~g sample were taken at intermediate temperatures to confirm the validity of the temperature dependent peak intensity measurement. Additional temperature dependent background data on the 55~mg sample were measured at $\textbf{Q}$ = (-0.504, 0.347, 0) and (-0.504, 0.417, 0).

\section*{Acknowledgements}
We acknowledge enlightening discussions with K. Lefmann and his group, and K. Beauvois and A. Piovano for assistance at the Institut Laue-Langevin, Grenoble. We thank the Institut Laue-Langevin for the allocated beamtime on IN22 (CRG-2807) and ThALES (5-41-1154), and the Paul Scherrer Institut for the beamtime at GPS. The project has received funding from the European Union’s Horizon 2020 research and innovation program under the Marie Skłodowska-Curie grant agreement No 884104 (PSI-FELLOW-III-3i). J.~K., Q.~W., and J.~C. acknowledge support from the Swiss National Science Foundation (200021$\_$188564). J.K. is further supported by the PhD fellowship from the German Academic Scholarship Foundation. N.B.C. thanks the Danish Agency for Science, Technology, and Innovation for funding the instrument center DanScatt and acknowledges support from the Q-MAT ESS Lighthouse initiative. M.M. is funded by the Swedish Research Council (VR) through a neutron project grant (Dnr. 2016-06955) as well as the KTH Materials Platform.

\section*{Author contributions}
D.G.M., J.C. initiated the project and planed it together with G.S., M.J., and N.B.C.. The sample was grown by M.O., N.M. and T.K., and prepared by J.K. and Q.W.. The uniaxial pressure cell was designed by D.B. and M.B. conducted the McStas simulations. G.S., M.B., F.B., C.N.W., M.M, Y.S., M.J. N.C.B, J.C. and D.G.M performed the experiments. The data were analyzed by G.S., J.Choi and D.G.M.. G.S., D.G.M., M.J. N.B.C. and J.C. wrote the paper with inputs from all co-authors.

\section*{Additional Information}
Correspondence and requests for materials should be addressed to G.S., J. C. or D.G.M
\section*{Competing financial interests}
The authors declare no competing interests.
\section*{Data availability}
The data that support the plots within this paper are available from the corresponding authors upon reasonable request. 

\bibliographystyle{naturemag}
\bibliography{LSCObib}

\end{document}